\documentclass[conference]{IEEEtran}
\usepackage{color,graphicx}
\usepackage{hyperref}
\usepackage{amsmath}
\usepackage{caption}
\usepackage{subcaption}
\usepackage{pbox}
\usepackage{epstopdf}
\usepackage{epsfig}
\definecolor{Gray}{gray}{0.9}
\usepackage[table]{xcolor}
\usepackage{xcolor}
\usepackage{tikz}

\ifCLASSINFOpdf
\else
\fi

\makeatletter

\newcommand{\Rmnum}[1]{\expandafter\@slowromancap\romannumeral #1@}
\makeatother
\begin{document}
\title{Hidden-nodes in coexisting LAA \& Wi-Fi: a measurement study of real deployments}
\author{Vanlin Sathya, Muhammad Iqbal Rochman, and Monisha Ghosh
\IEEEauthorblockN{}
\IEEEauthorblockA{University of Chicago, Chicago, Illinois-60637.
{Email: \{vanlin, muhiqbalcr, monisha\}@uchicago.edu}}
}
\maketitle

\begin{abstract}
LTE-Licensed Assisted Access (LAA) networks are beginning to be deployed widely in major metropolitan areas in the US in the unlicensed 5 GHz bands, which have existing dense deployments of Wi-Fi. This provides a real-world opportunity to study the problems due to hidden-node scenarios between LAA and Wi-Fi. The hidden node problem has been well studied in the context of overlapping Wi-Fi APs. However, when Wi-Fi coexists with LAA, the hidden node problem is exacerbated since LAA cannot use the well known Request-to-Send (RTS)/Clear-to-Send (CTS) mechanism to resolve contentions, resulting in throughput degradation for Wi-Fi. In this paper we describe detailed measurements and conclusions from experiments on the campus of the University of Chicago which presents a perfect hidden node scenario where Wi-Fi access points (APs) controlled by us and a LAA base-station (BS) deployed by AT\&T are hidden from each other, but the clients are not. We performed careful experiments in three different regions of the coexistence area: (i) clients midway between LAA \& Wi-Fi; (ii) clients close to the Wi-Fi AP; and (iii) clients close to the LAA BS. Our results show that in a situation where LAA uses an aggregate of three unlicensed channels (60 MHz bandwidth) which overlap with a 80 MHz Wi-Fi transmission, the Wi-Fi throughput at client devices suffers considerably. Overall, Wi-Fi performance is impacted by the hidden node problem more severely than LAA. In the best outdoor conditions, the throughput of  LAA and Wi-Fi is reduced by 35\% and 97\% respectively when coexisting with each other as compared when the other system is not present. Furthermore, we conclude that when both LAA and Wi-Fi use multiple 20 MHz channels, and there are multiple Wi-Fi APs coexisting with LAA on the same set of channels, the choice of Wi-Fi primary channels can have a significant impact on LAA throughput.

\end{abstract}


\section{Introduction}

LTE-Licensed Assisted Access (LAA), which uses Listen Before Talk (LBT) to access the 5 GHz unlicensed spectrum was specified by 3GPP Release-13 in 2016 \cite{3gpp}. While LBT is similar to its Wi-Fi counterpart, Carrier Sense Multiple Access with Collision Avoidance (CSMA-CA), the difference in access parameters like transmission duration (TXOP) and energy detection (ED) levels between LAA and Wi-Fi can lead to reduced throughput when the two systems coexist. While coexistence between LAA and Wi-Fi has been an active research area for a few years now, these studies are based primarily on theoretical analyses~\cite{TON,TCCN}, simulations~\cite{iqbal2017impact}, and limited experiments evaluating the effects of ED, TXOP, optimal resource scheduling, etc. With widespread deployments of LAA in many US cities, it is now possible to research coexistence performance of Wi-Fi and LAA in real-world scenarios. This research using real deployments can serve as a basis for improving coexistence of next-generation wireless networks in unlicensed spectrum~\cite{ACM} e.g. 5G NR-U and Wi-Fi 6E.

In our recent work~\cite{sathya2020measurement}, we measured the coexistence performance of overlapping LAA and Wi-Fi networks in various locations in Chicago. We believe that this was the first such measurement campaign that studied LAA and Wi-Fi performance in real deployments. This measurement campaign revealed several aspects of LAA that have not been adequately addressed in the research literature to date: 
(i) all LAA deployments we encountered aggregate three 20 MHz unlicensed channels, thus potentially creating increased interference to Wi-Fi operations;
(ii) the majority of Wi-Fi deployments today are 40 MHz and 80 MHz, while most of the previous literature focused only on a single LAA and a single Wi-Fi cell using the same 20 MHz channel;
(iii) multi-channel Wi-Fi and multi-channel LAA usage create coexistence scenarios that have not been adequately studied;
(iv) even though LAA deployments are primarily outdoors and Wi-Fi deployments are indoors, the signal strength of both at client devices operating outdoors is comparable, leading to increased coexistence and hidden node problems;
and (v) the impact of LAA on Wi-Fi latency and the larger TXOPs used by LAA need further study and analysis.

Since our previous measurement campaigns were conducted on the streets of downtown Chicago, we were limited by the deployed locations and number of Wi-Fi access points (APs), the number of Wi-Fi clients, and inability to use packet monitoring tools like Wireshark. In order to perform controlled experiments we needed a LAA base-station (BS) where we could control the number and location of coexisting Wi-Fi APs. Fortunately, We were able to identify such an area on the University of Chicago (UChicago) campus near the university bookstore where an AT\&T LAA BS was recently deployed. We deployed our own Wi-Fi APs inside the bookstore, on the same LAA channels and discovered that this deployment creates a hidden-node problem between the Wi-Fi APs and the LAA BS. While the hidden-node problem has been well studied in the context of traditional overlapping Wi-Fi APs in theory, simulations and experiments, we believe that the same rigorous study has not yet happened for hidden nodes between Wi-Fi and LAA, especially in the experimental context of this paper.
Our experiments lead to the conclusion that in a hidden-node scenario with LAA, Wi-Fi experiences a performance degradation that is more severe than that experienced by LAA, thus indicating a need to develop more advanced protocols for coexistence between LAA and Wi-Fi.

The paper is organized as follows. Section II provides a brief overview of LAA \& Wi-Fi medium access mechanisms, existing literature on LAA and Wi-Fi coexistence, including the Wi-Fi/LAA hidden-node scenario. Section III describes the experimental setup of the deployment we studied and the tools used to measure performance. Section IV describes the measurements and accompanying discussions and conclusions are presented in Section V.

\section{Overview of LAA and Wi-Fi Coexistence}

In this section we  summarize the main differences between LAA and Wi-Fi, followed by a discussion of existing research on LAA and Wi-Fi coexistence, both generally and specific to the hidden-node problem.

\subsection{Comparison of LAA and Wi-Fi}

LAA as specified by 3GPP in Release 13 \cite{lagen2019new}, permits downlink transmissions only in unlicensed channels and all uplink transmissions and signaling occur over the licensed channel. While the medium access protocol (MAC) is similar to Wi-Fi's, there are key differences as described below.

\subsubsection{Energy Detection Threshold}
Both LBT, as specified by LAA, and CSMA-CA, as specified by Wi-Fi, seek to avoid collisions by backing off from transmitting when the channel is busy. Both schemes define a busy channel as an energy level higher than a set threshold observed over the specified channel. LBT specifies an ED threshold of -72 dBm over 20 MHz, while Wi-Fi uses an ED threshold of -62 dBm over 20 MHz for non-Wi-Fi signals and a preamble detection threshold of -82 dBm for Wi-Fi. We have shown in ~\cite{iqbal2017impact} that this asymmetry leads to lower throughput performance of Wi-Fi. In a hidden-node scenario where the Wi-Fi AP and the LAA BS cannot detect each other, the Wi-Fi performance will deteriorate further since the commonly used request-to-send/clear-to-send (RTS/CTS) protocol cannot be used between the two.

\subsubsection{Medium Access Category}

LAA's LBT mechanism specifies a different transmission opportunity (TXOP) duration for different types of downlink traffic as shown in Table~\ref{ltedltxop}. Voice and video use a smaller TXOP to satisfy the Quality of Service (QoS) requirements, which can be met by the use of short packets. However, for data, LAA specifies a maximum TXOP of 8 ms to maximize throughput, or 10 ms if there is no coexisting Wi-Fi. In most realistic traffic scenarios we measured, such as Data, Data + Video, Data + Streaming, we observe that the LAA BS uses a maximum TXOP of 8 ms. 

Similarly, Wi-Fi also defines access categories for different types of traffic as shown in Table~\ref{wtxop} for 802.11ac. Compared to LAA, the data traffic (Best Effort and Background) is allocated a smaller TXOP duration of 2.528 ms, compared to 8 ms used by LAA. This asymmetry with respect to LAA can lead to LAA occupying the channel longer than Wi-Fi on average.
 
Additionally, both access categories also define initial Clear Channel Assessment (CCA) duration, and minimum and maximum contention window (CW) sizes. The initial CCA is used to determine whether the channel is clear for transmission at the beginning by observing it for a set period. For Wi-Fi, this period is defined as Arbitrary Inter-Frame Spacing (AIFS). For LAA, the CW is exponentially increased when the Hybrid Automatic Repeat Request - Acknowledgment (HARQ-ACK) reply contains 80\% Negative ACKs (NACKs), while for Wi-Fi, the CW is exponentially increased when a single transmission fails (no ACK received).

\begin{table}[htb]
\centering
 \caption{Access Categories in Downlink LAA~\cite{sathya2020measurement}}
     \begin{tabular}{|p{55pt}|p{20pt}|l|l|l|}
\hline
\cellcolor{Gray} \textbf{Access Class \# (DL)} & \cellcolor{Gray} \textbf{Initial CCA} & \cellcolor{Gray} \textbf{CWmin} & \cellcolor{Gray} \textbf{CWmax} & \cellcolor{Gray} \textbf{$TXOP$} \\
\hline\hline
1 (Voice) & 25 $\mu$ s & 30 $\mu$ s & 70 $\mu$ s & 2 ms \\
\hline
2 (Video)& 25 $\mu$ s & 70 $\mu$ s & 150 $\mu$ s & 3 ms \\
\hline
3 (Best Effort)& 43 $\mu$ s & 150 $\mu$ s & 630 $\mu$ s & 8 ms or 10 ms \\
\hline
4 (Background)& 79 $\mu$ s & 150 $\mu$ s & 10.23 ms & 8 ms or 10 ms \\
\hline
\end{tabular}
\label{ltedltxop}
\end{table}

\begin{table}[htb]
\centering
 \caption{Access Categories in Wi-Fi (802.11ac)~\cite{sathya2020measurement}}
     \begin{tabular}{|p{70pt}|p{25pt}|l|l|l|}
\hline
\cellcolor{Gray} \textbf{Access Category} & \cellcolor{Gray} \textbf{AIFS} & \cellcolor{Gray} \textbf{CWmin} & \cellcolor{Gray} \textbf{CWmax} & \cellcolor{Gray} \textbf{$TXOP$} \\
\hline\hline
Voice (AC\_VO) & 18 $\mu$ s & 27 $\mu$ s  & 63 $\mu$ s & 2.08 ms \\
\hline
Video (AC\_VI) & 18 $\mu$ s & 62 $\mu$ s & 135 $\mu$ s & 4.096 ms \\
\hline
Best Effort (AC\_BE) & 27 $\mu$ s & 135 $\mu$ s & 9.207 ms & 2.528 ms \\
\hline
Background (AC\_BK) & 63 $\mu$ s & 135 $\mu$ s & 9.207 ms & 2.528 ms \\
\hline
\end{tabular}
\label{wtxop} -
\end{table}

\subsection{Related Work on LAA and Wi-Fi Coexistence}

There has been extensive research on coexistence of Wi-Fi and LTE, mainly focused on using analysis and simulation to study fair coexistence. In \cite{7460494}, the authors explored design of LBT schemes for LAA as a means of providing equal opportunity channel access in the presence of Wi-Fi. Similarly, \cite{tao2015enhanced} proposed an enhanced LBT algorithm with contention window size adaptation for LAA to achieve fair channel access and quality of Service (QoS) fairness.
Comprehensive experiments on the coexistence of LAA and Wi-Fi are described in \cite{jian2015coexistence, QC}, without addressing the hidden-node problem. In our prior work, \cite{TON, iqbal2017impact} we studied the impact of different sensing duration's and ED thresholds via analysis, simulations and limited experiments using a software-defined-radio (SDR) testbed and concluded that the overall throughput of both coexisting systems improved if both Wi-Fi and LTE employed an ED threshold of -82 dBm..

\begin{figure}[!tb]
 \centering
 \includegraphics[scale=0.5]{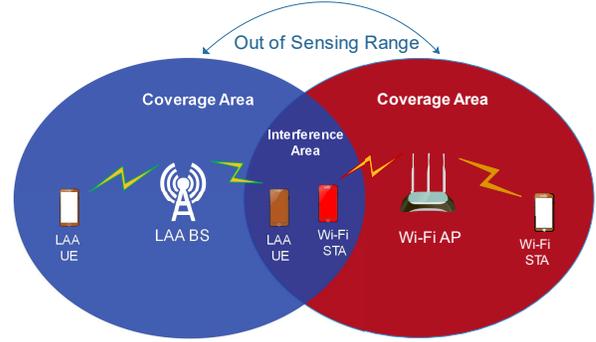}
 \caption{Hidden Node Problem in LAA Wi-Fi Coexistence.}
 \label{hidden}
\end{figure}

\subsection{Hidden-nodes in LAA and Wi-Fi Coexistence}

The hidden-node problem is a common one in wireless networking, where interference occurs at a receiver because the transmitter is "hidden" from an interfering transmitter. In traditional Wi-Fi/Wi-Fi coexistence, or overlapping service areas, the hidden node problem is solved by the use of the RTS/CTS packets which are received by all radio transceivers in range. However, since LAA and Wi-Fi cannot decode each others packets, this solution does not work when the two coexist in an overlapping coverage area.
This is shown in Fig.~\ref{hidden}, where the LAA BS is not aware of the Wi-Fi APs transmission in the interference zone. Furthermore, the asymmetry of LAA and Wi-Fi energy detection levels may cause the Wi-Fi client's RTS packet to be ignored by the BS. In~\cite{campos2020analysis}, this hidden-node problem is addressed, in 20 MHz, by ns-3 simulations. The Channel Quality Indicator (CQI), Reference Signal Received Power (RSRP), and Reference Signal Received Quality (RSRQ) metrics at the LAA client device are used to develop a collision detection algorithm that is then used to modify the LAA MAC parameters. Other papers mostly address the hidden-node problem in the context of LTE-Unlicensed (LTE-U), which is a variant of LTE unlicensed channel access that uses a duty cycle method instead of LBT~\cite{baswade2018lte, baswade2018law}. Hence, there is no research on the hidden-node issue of LAA \& Wi-Fi coexistence that is based on analyzing real deployments which use multiple 20 MHz channels for both Wi-Fi and LAA. We seek to address this gap in this paper.

\section{Experiment Set-up and Tools}

There are a number of Release 13 compliant LAA base-stations deployed on the UChicago campus by AT\&T mostly on lamp-posts near campus buildings. We chose a particular deployment for this study where a LAA BS is mounted on a pole near the UChicago bookstore. The BS aggregates up to three Wi-Fi channels: 149, 153, and  157 in the U-NII 3 band and has the capability of 2 $\times$ 2 MIMO transmissions with a maximum modulation coding scheme (MCS) of 256 QAM. In this section we describe the set-up and tools used in the experiment.

\begin{figure*}[ht]
\begin{subfigure}{.33\textwidth}
  \centering
 \includegraphics[width=5.2cm]{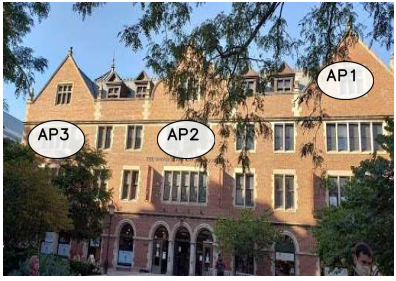}
  \caption{Wi-Fi AP Deployment}\label{e1}
\end{subfigure}
\begin{subfigure}{.23\textwidth}
  \centering
    \includegraphics[width=3.2cm]{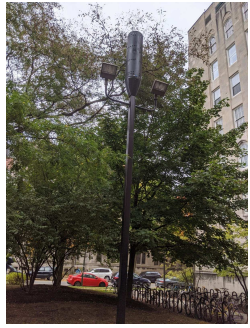}
  \caption{LAA BS Deployment}\label{e2}
\end{subfigure}
\begin{subfigure}{.43\textwidth}
  \centering
   \includegraphics[width=7.2cm]{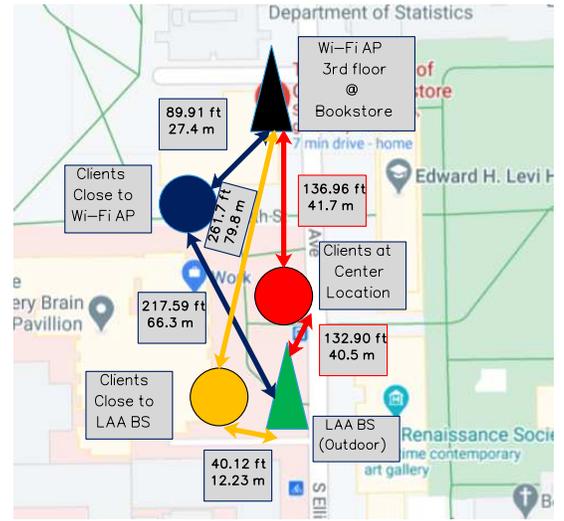}
  \caption{Experiment Location}\label{e3}
\end{subfigure}

\caption{UChicago Bookstore Experiment Location}
\label{e4}
\end{figure*}

\begin{figure*}[ht]
\begin{subfigure}{.33\textwidth}
  \centering
 \includegraphics[width=5.2cm]{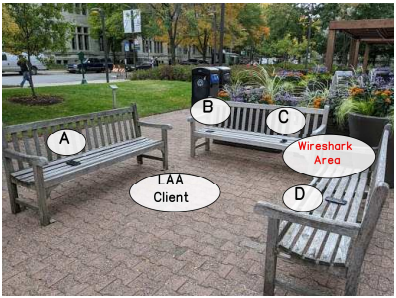}
  \caption{Clients at Center (S1)}\label{e5}
\end{subfigure}
\begin{subfigure}{.33\textwidth}
  \centering
    \includegraphics[width=5.2cm]{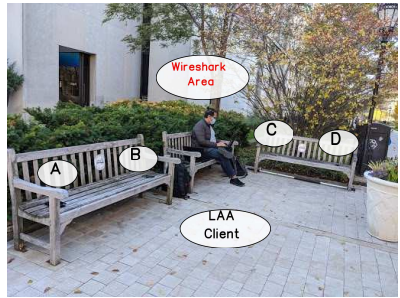}
  \caption{Clients Close to Wi-Fi AP (S2)}\label{e6}
\end{subfigure}
\begin{subfigure}{.33\textwidth}
  \centering
   \includegraphics[width=5.2cm]{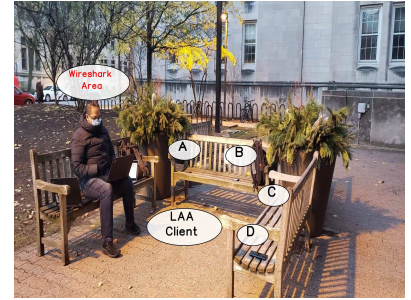}
  \caption{Clients Close to LAA (S3)}\label{e7}
\end{subfigure}

\caption{Deployment of LAA by AT\&T, and coexisting Wi-Fi on Channels 149, 153 and 157}
\label{e8}
\end{figure*}

\subsection{Measurement Tools: SigCap, Network Signal Guru (NSG), and Wireshark}

We use SigCap, an Android app we developed that extracts information from the radios on a client device without requiring root access \cite{sathya2020measurement}. SigCap collects data from the cellular receiver (RSRP, RSSI, LTE band, etc.) and Wi-Fi receiver (RSSI, channel number, etc.) which can be easily exported for further analysis. However, to obtain more detailed measurements, we also use the Network Signal Guru (NSG) app \cite{NSG} which uses root access to extract additional LTE information such as throughput, latency, TXOP, signal-to-interference-and-noise ratio (SINR), block error rate (BLER), and allocated resource blocks (RB) for each channel. Wireshark was used to capture additional Wi-Fi information such as Wi-Fi amendment (e.g., 802.11ac), number of clients on each channel, number of streams used for transmission, modulation coding schemes, and throughput.

\subsection{Location of Experiments}
We deployed three Wi-Fi APs inside the bookstore, near windows facing the LAA BS as shown in Fig.~\ref{e4} (a). AP1 is on the fourth floor of the building, and AP2 \& AP3 are on the third floor. All distances shown are measured on the ground, not point-to-point.
Fig.~\ref{e4} (b) shows the pole where the LAA BS is mounted in the black enclosure at the top. We deduce that this is indeed the BS location from the high SINR and RSRP value recorded by NSG near the pole. 
Fig.~\ref{e4} (c) shows the diagram of the experiment location. We deployed the clients in three locations shown as red, blue, and yellow circles representing the following scenarios:
\begin{itemize}
\item \textbf{Scenario S1 (red): clients midway between LAA BS \& Wi-Fi APs.} In this scenario, the Wi-Fi and LAA clients are midway between the LAA BS and the Wi-Fi APs at a distance of 41.7 m from the Wi-Fi APs and 40.5 m from the LAA BS, as shown in Fig.~\ref{e8} (a). 
\item \textbf{Scenario S2 (blue): clients close to Wi-Fi APs.} All clients are closer to the APs at a distance of 27.4 m from the Wi-Fi APs and 66.3 m from the LAA BS, as shown in Fig.~\ref{e8} (b).
\item \textbf{Scenario S3 (yellow): clients close to the LAA BS.} All clients are closer to the LAA BS at a distance of 79.8 m from the Wi-Fi APs and 12.23 m from the LAA BS, as shown in Fig.~\ref{e8} (c).
\end{itemize}
The Wi-Fi clients are placed on benches about 2 ft above ground, the LAA client is hand-held at about 5 ft above the ground, and the Wi-Fi APs and LAA BS are at a height of approximately 50 ft and 30 ft, respectively.

\subsection{Devices and Parameters used in the experiment}
Table~\ref{exp8} shows the various devices used in the experiments. All three APs operate on Channel 155 using the 80 MHz mode with each AP on a different 20 MHz primary channel as shown in Table~\ref{exp8}. We disabled dynamic frequency selection to prevent the APs from moving to other channels. We also used Wireshark to verify that the only Wi-Fi data transmissions that took place during the duration of our experiments were from the APs we deployed, even though we detected the presence of beacons from other campus APs. Thus, our experiment included only the 3 APs and the LAA BS coexisting with each other.
We use a total of 5 phones as client devices. For the Wi-Fi-only experiments, all 5 are used as Wi-Fi clients and for the Wi-Fi/LAA coexistence experiments device A, B, C, \& D in Table~\ref{exp8}) are used as Wi-Fi clients and device E, a Google Pixel 3 phone, is used as the LAA client. All client devices are equipped with NSG and SigCap.

\begin{table}[!t]
\scriptsize
\caption{Devices and configurations}
\centering
\begin{tabular}{|c|c|}
\hline
\bfseries Parameter &\bfseries Value \\ [0.4ex]
\hline\hline
Wireshark Laptop & 3 \\
\hline
Wi-Fi AP & \pbox{30cm}{ \vspace{0.2cm} AP1 $\rightarrow$ Netgear (Primary Ch. 149) \\ AP2 $\rightarrow$ TP-Link (Primary Ch. 157) \\ AP3 $\rightarrow$ Netgear (Primary Ch. 153) \\ }\\
\hline
Wi-Fi Device Clients &   \pbox{30cm}{ \vspace{0.2cm} A $\rightarrow$ T-Mobile Google Pixel 3 \\ B $\rightarrow$ Motorola Edge+ \\ C $\rightarrow$ Samsung S9\\ D $\rightarrow$ Xiaomi \\ E $\rightarrow$ AT\&T Google Pixel 3 \\ } \\
\hline
LAA Device Client &  E $\rightarrow$ AT\&T Google Pixel 3 \\
\hline
802.llac 80 MHz supported clients & A, B, C, E \\
\hline
802.lln 40 MHz supported client & D \\
\hline
\end{tabular}
\label{exp8}
\end{table}

\subsection{Verifying the hidden-node}
In order to verify that this deployment was indeed a hidden-node, we used SigCap to measure the RSSI of the LAA BS at each of the APs, and the RSSI of the Wi-Fi APs at the LAA BS (using a phone mounted on a 30 ft pole). Table~\ref{tableRssi} shows that the strongest Wi-Fi signal level at the LAA BS is from AP2 at -77 dBm, and the strongest LAA BS's signal level is at AP2 at -84 dBm. These signal levels are well below the ED thresholds of -72 dBm used by LAA and -62 dBm used by Wi-Fi. Hence, we are confident that the LAA BS and the Wi-Fi APs will not attempt to back-off to each other.

\begin{table}[!t]
\caption{RSSI Measurement of LAA BS and Wi-Fi APs}
\centering
\begin{tabular}{|l|c|c|c|}
\hline
& \bfseries AP1 & \bfseries AP2 & \bfseries AP3 \\
\hline
\bfseries RSSI of LAA BS at AP$n$ & -88 dBm & -84 dBm & N/A\\
\hline
\bfseries RSSI of AP$n$ at LAA BS & -80 dBm & -77 dBm & -81 dBm \\
\hline
\end{tabular}
\label{tableRssi}
\end{table}

\subsection{Traffic Types}
We conducted our experiments using the following traffic types:
\begin{itemize}
    \item \textbf{Data (D):} Pure data traffic assumed as a full buffer, generated by downloading a large YUV dataset ($>$10 GB) from Derf Test Media Collection \cite{YUV}.
    \item \textbf{Video (V):} A Youtube video is downloaded, with a resolution of 1920$\times$1080 and bit-rate of 12 Mbps.
    \item \textbf{Data + Video (D+V):} Combination of data and video traffic as described above.
    \item \textbf{Streaming (S):} A live stream video on Youtube is loaded, with a resolution of 1280$\times$720 and a bit rate of 7.5 Mbps. 
    \item \textbf{Data + Streaming (D+S):} Combination of data and streaming traffic as described above.
\end{itemize}

We observed (using NSG) that 95\% - 100\% of the available LAA resource blocks were always allocated to our client device: this is possibly due to (a) COVID-19, as there were very few people on campus and (b) LAA is only available on newer phones that tend to be more expensive and hence are not yet widely used. This allows us to perform a controlled coexistence experiment as follows:
\begin{itemize}
    \item \textbf{Wi-Fi/Wi-Fi Scenario:} All five clients are connected to a Wi-Fi AP and initiate downloads of the same traffic type. We assume in this case that no other device is associated with the LAA BS, which can be assumed to be not transmitting any data other than reference signals.
    \item \textbf{Wi-Fi/LAA Scenario:} Four clients are connected to Wi-Fi, and one client is connected to the LAA BS, all initiating downloads of the same traffic type.
\end{itemize}

\begin{table}[htb]
\scriptsize
 \caption{Baseline throughput of LAA alone (no Wi-Fi)}
     \begin{tabular}{|p{1.4cm}|p{1.1cm}|p{1.1cm}|p{0.8cm}|p{1.1cm}|p{1cm}|}
\hline
\cellcolor{Gray} \textbf{Scenarios} & \cellcolor{Gray} \textbf{D} & \cellcolor{Gray} \textbf{D+V} & \cellcolor{Gray} \textbf{S} & \cellcolor{Gray} \textbf{D+S} & \cellcolor{Gray} \textbf{V} \\
\hline\hline
S2: Licensed & 26.3 Mbps & 32.7 Mbps & 8.4 Mbps & 12.2 Mbps & 10.9 Mbps \\
\hline
S2: LAA 149 & 53.5 Mbps & 58.6 Mbps & - & 44.1 Mbps & 8.1 Mbps \\
\hline
S2: LAA 153  & 60.4 Mbps & 67.1 Mbps & - & 52.9 Mbps & 9.8 Mbps \\
\hline
S2: LAA 157 & 61.5 Mbps & 61.9 Mbps & - & 49.4 Mbps & 8.3 Mbps \\
\hline
\end{tabular}
\label{ta4}
\end{table}

\section{Representative measurements of LAA \& Wi-Fi}\label{is}

We use NSG to measure the TXOP, SINR, RBs, and throughput of LAA, with all results averaged over ten measurements in the same location. Similarly, Wireshark is used to measure Wi-Fi throughput.
The LAA BS uses a maximum of three unlicensed 20 MHz channels and a single 15 MHz licensed channel for a maximum total bandwidth of 75 MHz and 375 RBs.

\begin{figure*}[htb]
\begin{subfigure}{.33\textwidth}
  \centering
 \includegraphics[width=5.2cm]{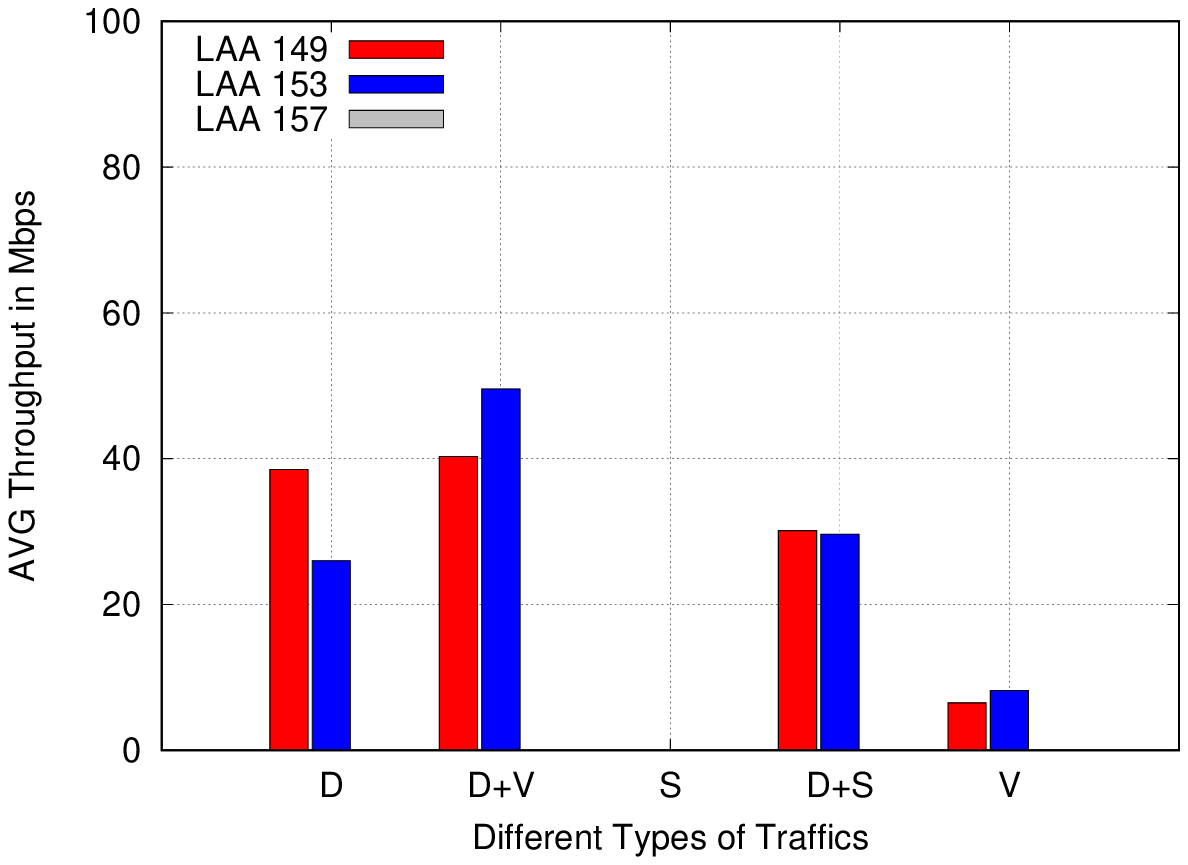}
  \caption{Clients at Center (S1)}
\end{subfigure}
\begin{subfigure}{.33\textwidth}
  \centering
    \includegraphics[width=5.2cm]{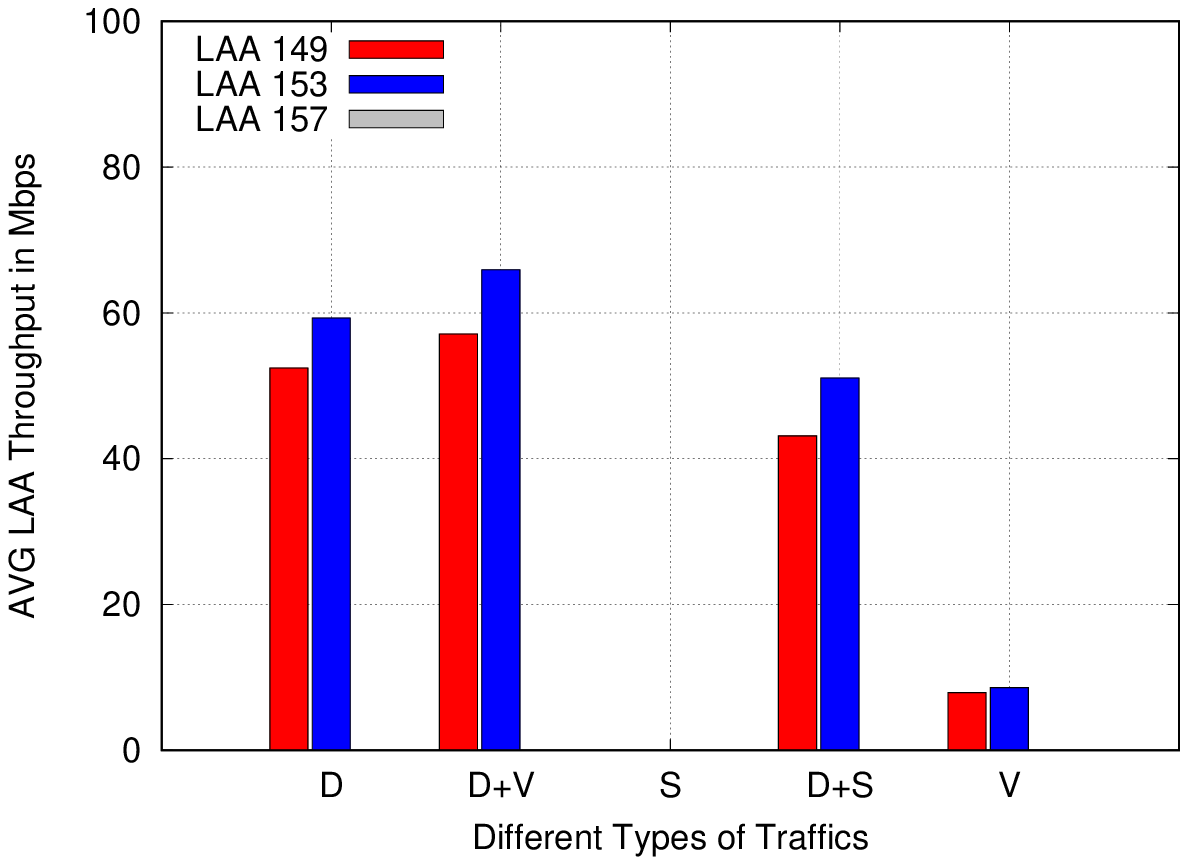}
  \caption{Clients Close to Wi-Fi APs (S2)}
\end{subfigure}
\begin{subfigure}{.33\textwidth}
  \centering
   \includegraphics[width=5.2cm]{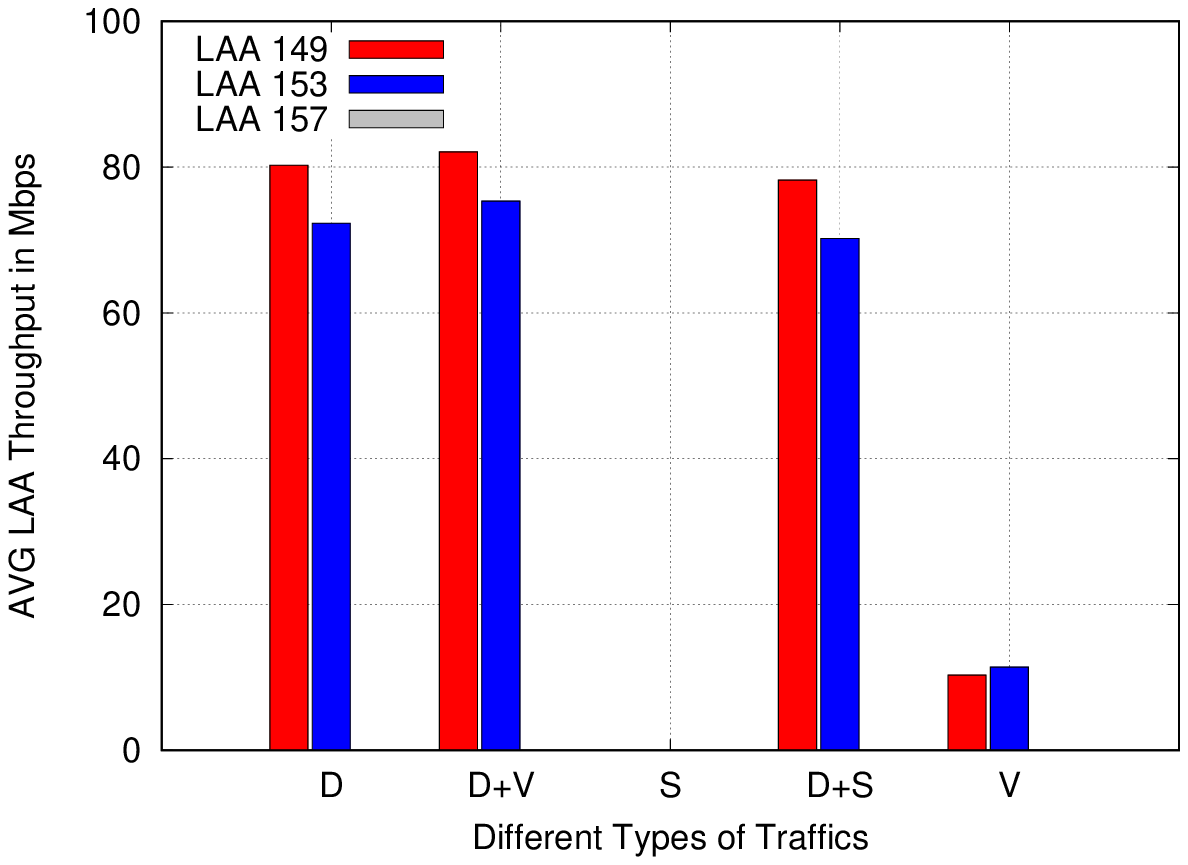}
  \caption{Clients Close to LAA BS (S3)}
\end{subfigure}
\caption{LAA throughput, Wi-Fi/LAA coexistence, all Wi-Fi clients associated to AP2}
\label{atla}
\end{figure*}

\subsection{Baseline LAA performance, Scenario S2}
We established a baseline LAA performance with Scenario S2, where all Wi-Fi clients were associated with the Wi-Fi APs without initiating any traffic: 1 client @ AP1, two clients @ AP2, 1 client @ AP3. The LAA client then initiated traffic according to the scenarios (D, D+V, S, D+S, V), and Table~\ref{ta4} shows the average throughput of LAA on each carrier and traffic type. We consistently measured an aggregate throughput of greater than 150 Mbps over the three unlicensed carriers, compared to about 25 Mbps over the single licensed carrier. Since the Wi-Fi clients are not engaged in data transmission, we measure a throughput of about 43 kbps, due to management frame transmissions alone.

\subsection{Measurements results, all scenarios}
AP2 was the only AP that the Wi-Fi clients could connect to in all the three scenarios. Hence, in this section, we present results with all Wi-Fi clients connected to AP2. In each scenario, the Wi-Fi and LAA clients were close to each other.

\subsubsection{LAA SINR and throughput when coexisting with Wi-Fi}
Table~\ref{laas} shows the average LAA SINR on Channels 149, 153, and 157. In all scenarios, the LAA SINR on Channel 157 is much lower compared to the other channels for all traffic types. This is because all Wi-Fi clients are associated with AP2 which uses Channel 157 as its primary channel. Since the default mode of the APs was set to use RTS/CTS, the traffic on the primary channel, Channel 157, is higher than on the other channels, even though the actual data transmission occurs over all channels. This leads to a reduced LAA SINR on Channel 157.

Fig.~\ref{atla} shows the LAA throughput in the different scenarios. There are two conclusions from this result: (i) when the traffic type is real-time video streaming, the unlicensed channels are not used at all. This is because the QoS requirements of real-time traffic cannot be guaranteed in the unlicensed band, and (ii) Channel 157 is not used by LAA at all. This could be either due to the increased management traffic from the clients that is sensed at the LAA BS or the reduced SINR value that is fed back to the LAA BS which uses the information to not allocate this channel for transmission. We also observe that the maximum TXOP of 8 ms is used on Channels 149 and 153 for D, D+V, and D+S, and TXOP of 3 ms is used for video (V) transmission and the average number of RBs allocated on these 2 channels is always greater than 90 each.

\begin{table}[htb]
 \caption{Average LAA SINR in Scenarios S1, S2 and S3}
     \begin{tabular}{|p{1.7cm}|p{1cm}|p{0.9cm}|p{0.5cm}|p{1cm}|p{0.9cm}|}
\hline
\cellcolor{Gray} \textbf{Scenarios} & \cellcolor{Gray} \textbf{D} & \cellcolor{Gray} \textbf{D+V} & \cellcolor{Gray} \textbf{S} & \cellcolor{Gray} \textbf{D+S} & \cellcolor{Gray} \textbf{V} \\
\hline\hline
S1: LAA 149 & 12 dB & 10 dB & - & 9 dB & 12.2 dB \\
\hline
S1: LAA 153  & 8 dB & 13 dB & - & 8 dB & 13 dB \\
\hline
\textcolor{red}{S1: LAA 157} & -3 dB & -1 dB & - & -0.5 dB & -4 dB \\
\hline
S2: LAA 149 & 14 dB & 12.6 dB & - & 9.7 dB & 12.8 dB \\
\hline
S2: LAA 153 & 10 dB & 11.4 dB & - & 8.2 dB & 14.2 dB \\
\hline
\textcolor{red}{S2: LAA 157} & -2 dB & -3.2 dB & - & -1.9 dB & -2.9 dB \\
\hline
S3: LAA 149 & 20.3 dB & 20.6 dB & - & 20 dB & 22 dB \\
\hline
S3: LAA 153 & 22.6 dB & 23 dB & - & 23.2 dB & 22.8 dB \\
\hline
\textcolor{red}{S3: LAA 157} & -2.1 dB & -3.4 dB & - & -4 dB & -3.6 dB \\
\hline
\end{tabular}
\label{laas}
\end{table}

Fig.~\ref{atla} also shows that the LAA throughput in Scenario S2 is higher than in Scenario S1, even though S2 is farther from the LAA BS than S1. This is because location S1 did not have direct line-of-sight to the LAA BS even though it was closer, due to tall trees in between. This is also evident from Table~\ref{laas} where the SINR in location S1 is consistently lower than in S2. Throughput, and SINR, are highest in S3 due to proximity and line-of-sight to the LAA BS. Comparing the results with the baseline LAA performance measured earlier at S2 on data (D) traffic, we observe a reduction of 61.2 Mbps in aggregate throughput from the baseline, which is around 35\% of baseline.

\begin{figure*}[ht]
\begin{subfigure}{.33\textwidth}
  \centering
  \includegraphics[width=5.2cm]{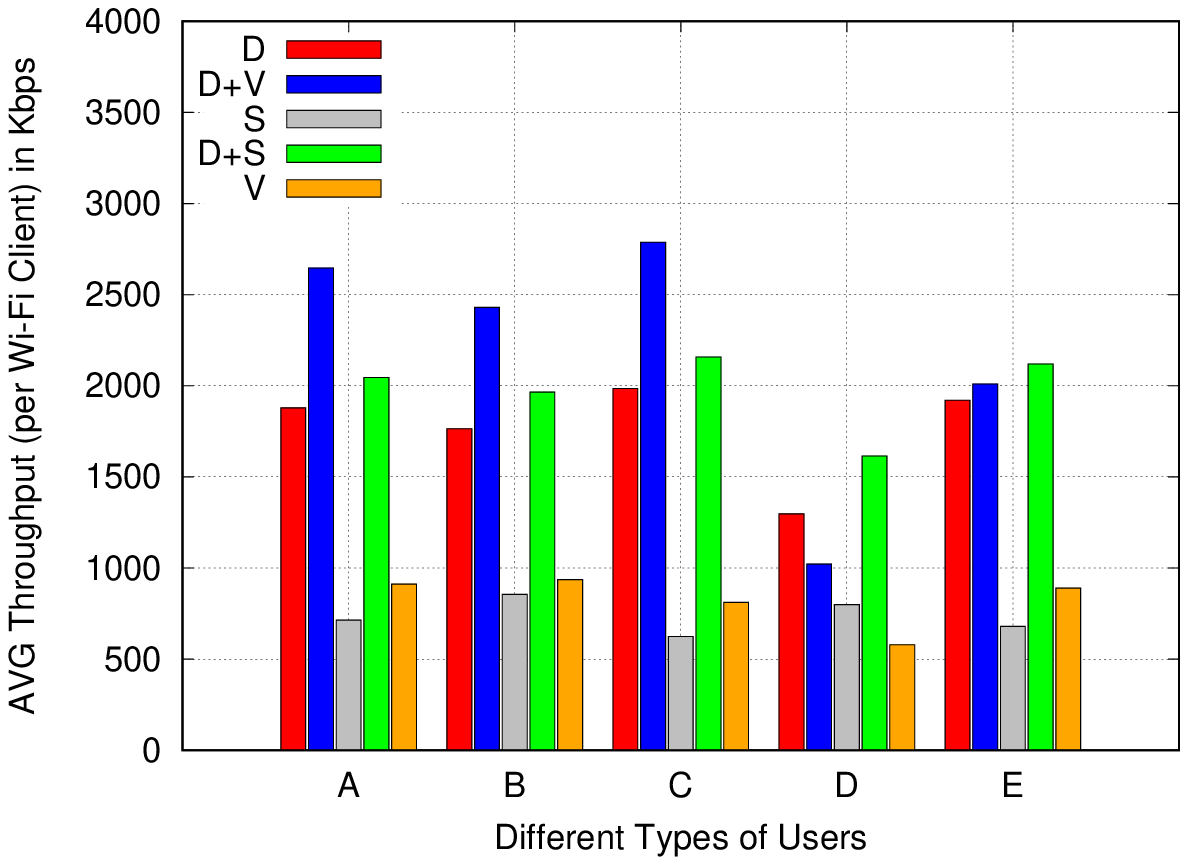}
  \caption{Clients at center (S1)}
\end{subfigure}
\begin{subfigure}{.33\textwidth}
  \centering
  \includegraphics[width=5.2cm]{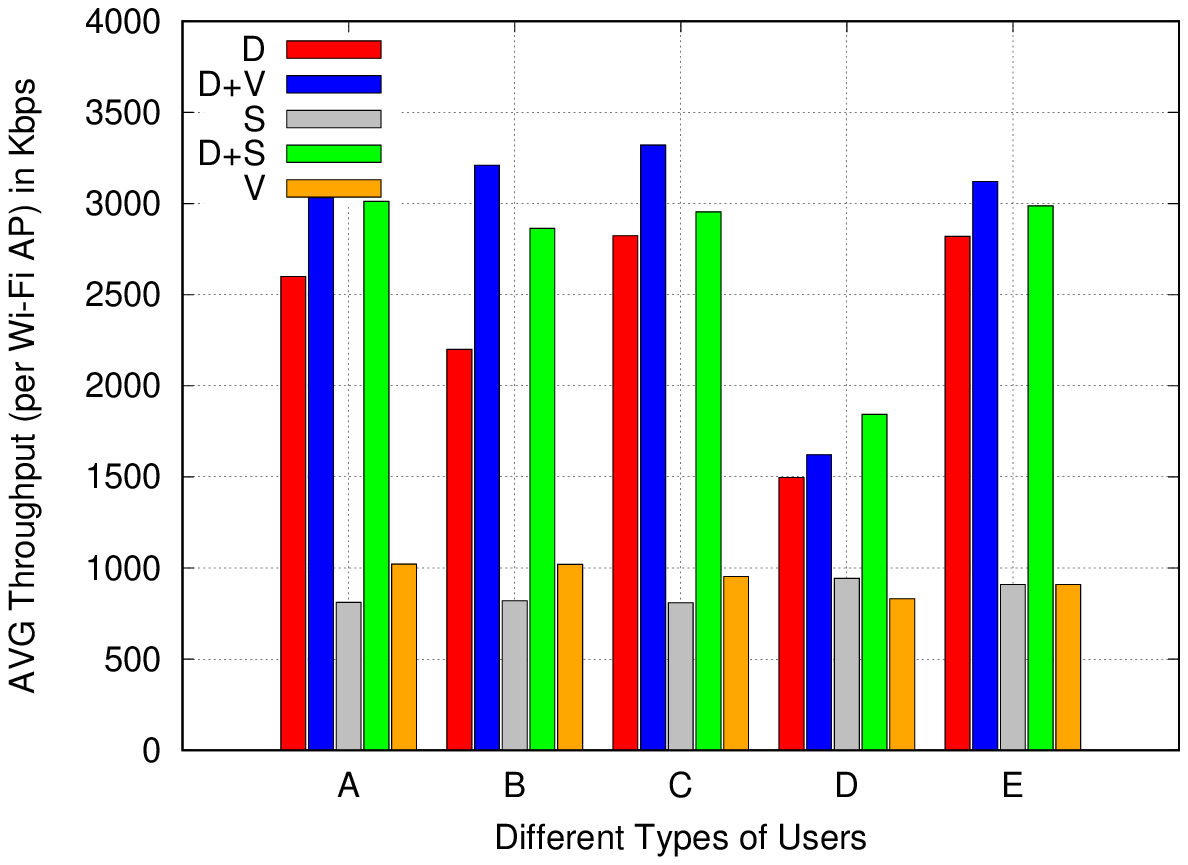}
  \caption{Clients close to Wi-Fi APs (S2)}
\end{subfigure}
\begin{subfigure}{.33\textwidth}
  \centering
  \includegraphics[width=5.2cm]{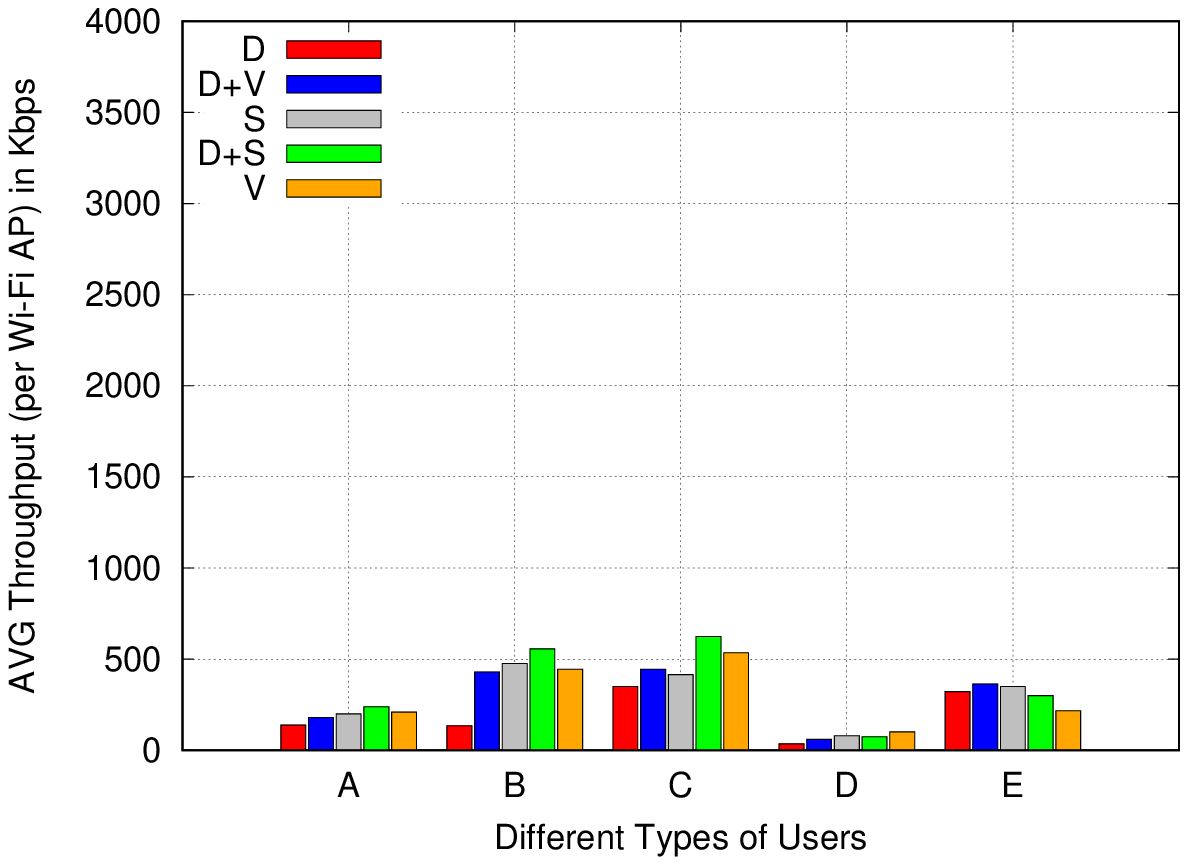}
  \caption{Clients close to LAA BS (S3)}
\end{subfigure}
\caption{Wi-Fi Throughput, Wi-Fi/Wi-Fi coexistence, all Wi-Fi clients associated to AP2}
\label{wifiTputW}
\end{figure*}

\begin{figure*}[ht]
\begin{subfigure}{.33\textwidth}
  \centering
  \includegraphics[width=5.2cm]{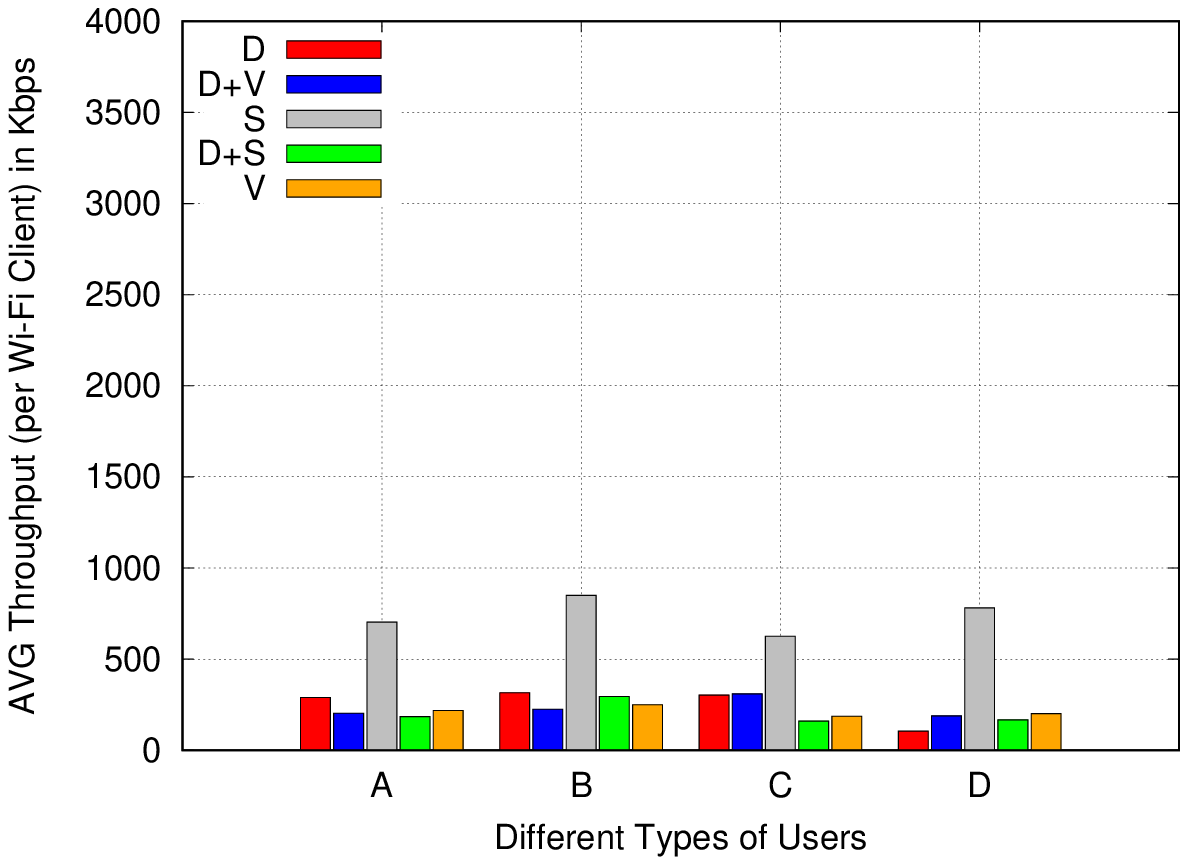}
  \caption{Clients at center (S1)}
\end{subfigure}
\begin{subfigure}{.33\textwidth}
  \centering
  \includegraphics[width=5.2cm]{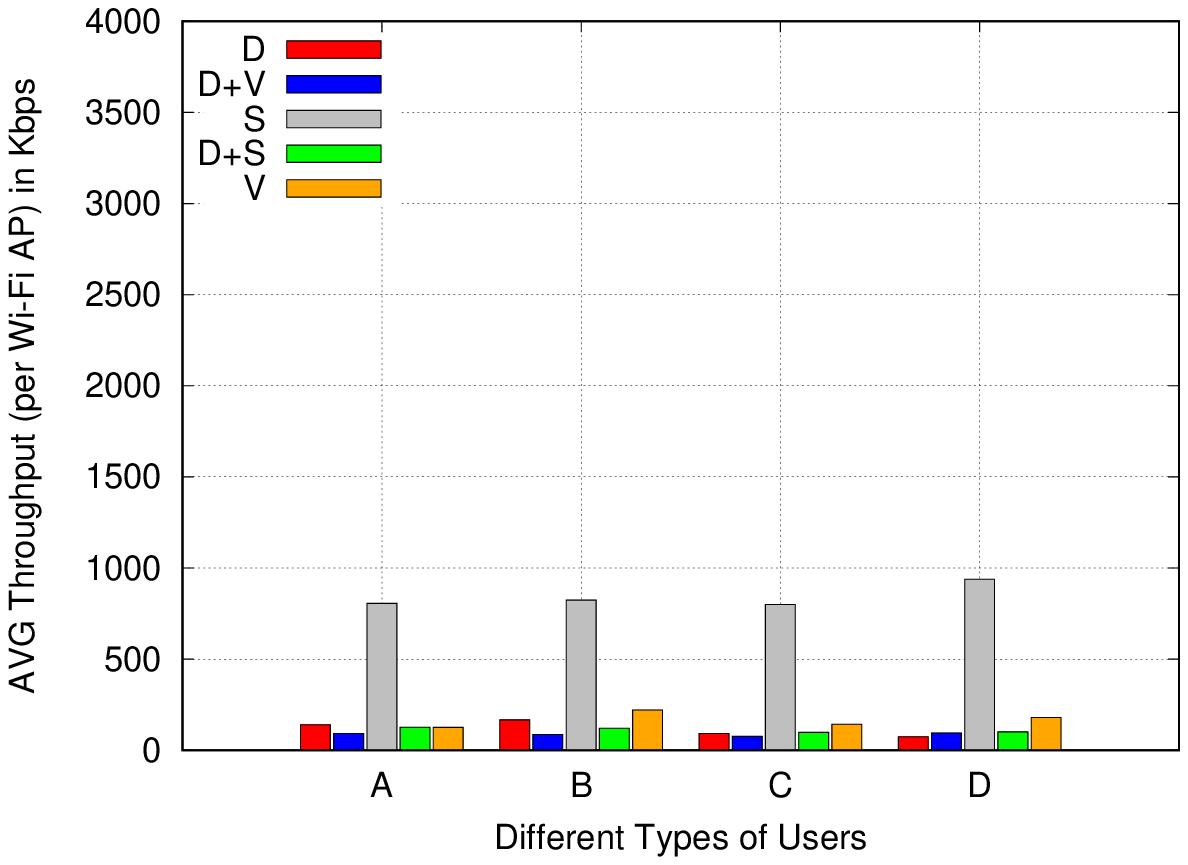}
  \caption{Clients close to Wi-Fi APs (S2)}
\end{subfigure}
\begin{subfigure}{.33\textwidth}
  \centering
  \includegraphics[width=5.2cm]{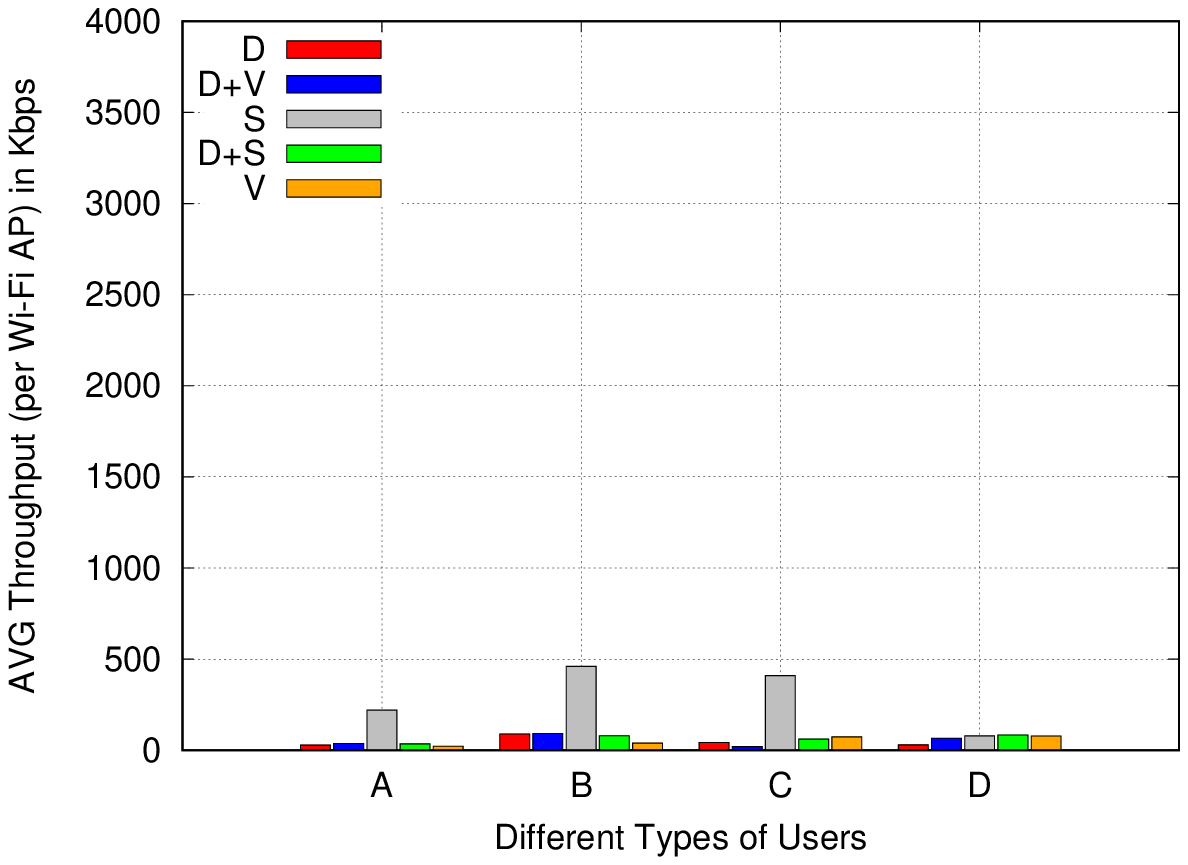}
  \caption{Clients close to LAA BS (S3)}
\end{subfigure}
\caption{Wi-Fi Throughput, Wi-Fi/LAA coexistence, all Wi-Fi clients associated to AP2}
\label{wifiTputL}
\end{figure*}

\begin{figure*}[ht]
\begin{subfigure}{.33\textwidth}
  \centering
 \includegraphics[width=5.2cm]{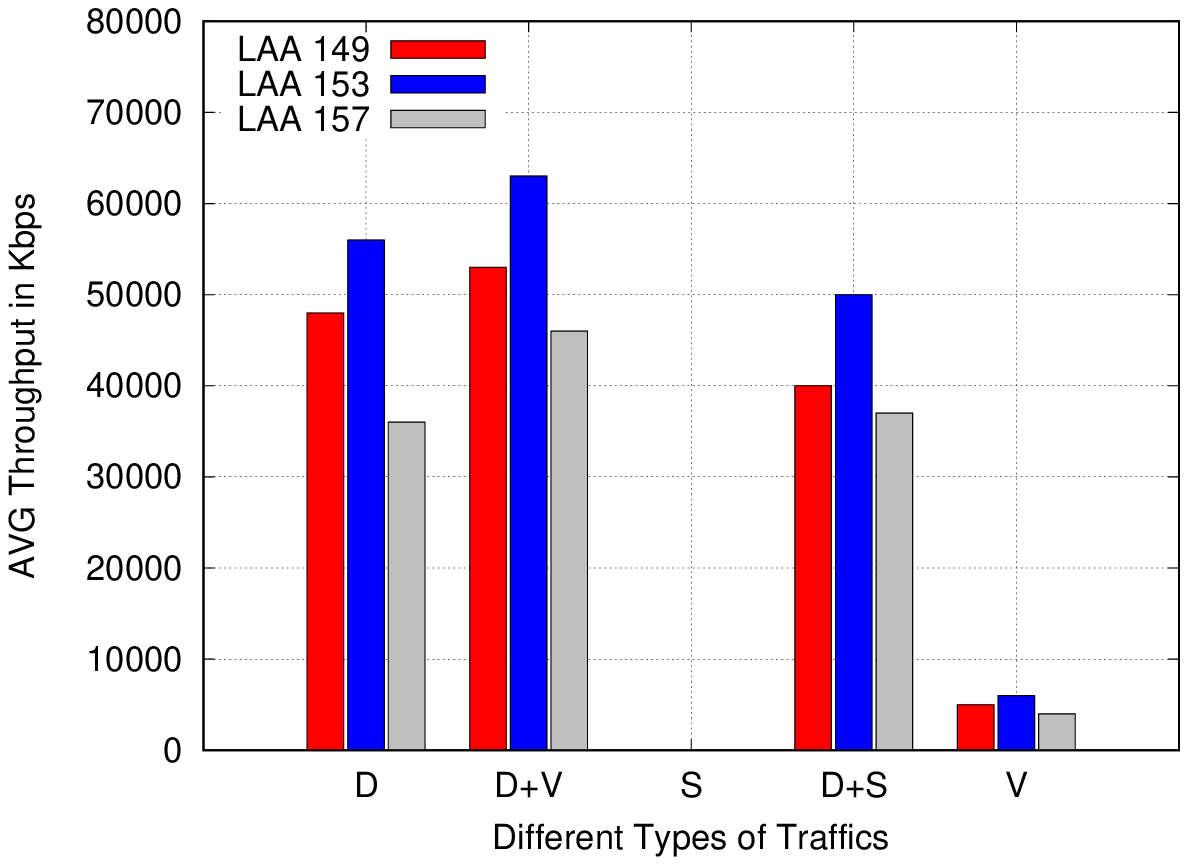}
  \caption{LAA Tput on Wi-Fi/LAA}\label{fig:laa36}
\end{subfigure}
\begin{subfigure}{.33\textwidth}
  \centering
    \includegraphics[width=5.2cm]{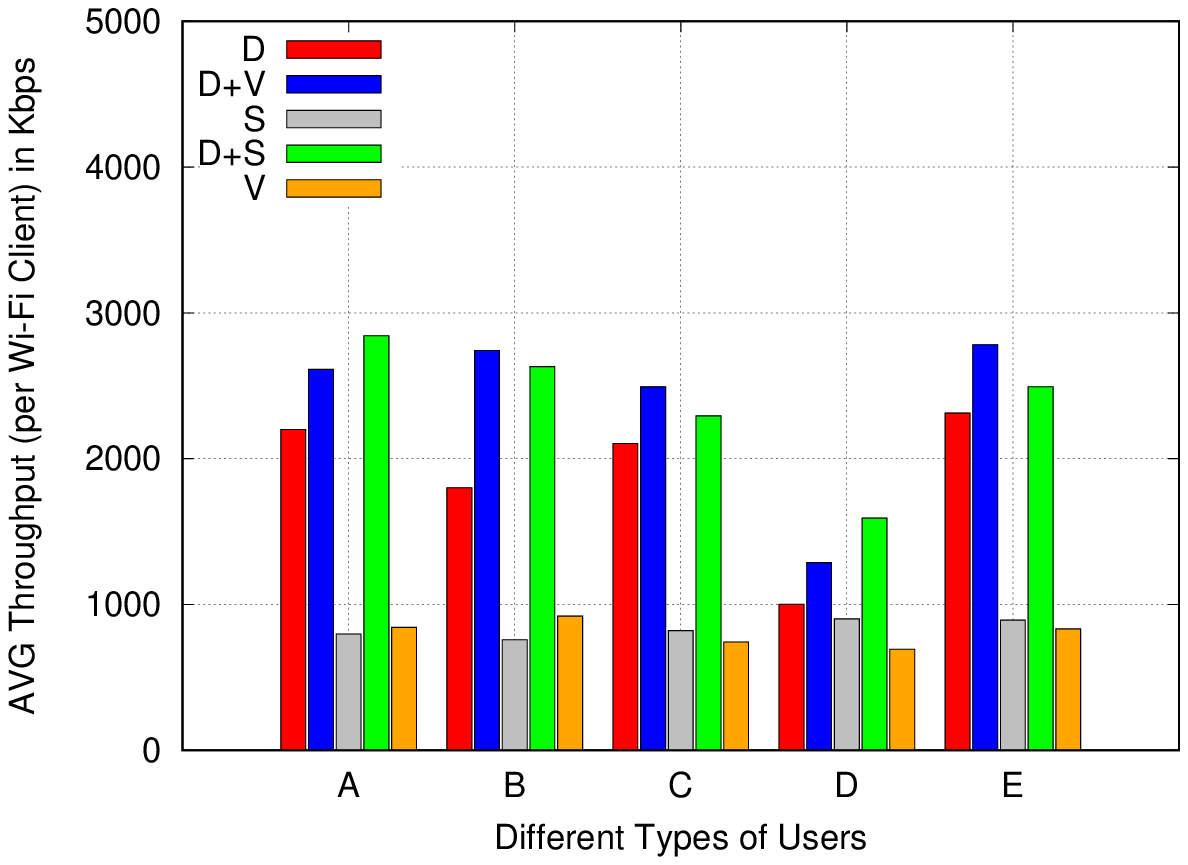}
  \caption{Wi-Fi Tput on Wi-Fi/Wi-Fi}\label{fig:laa149}
\end{subfigure}
\begin{subfigure}{.33\textwidth}
  \centering
   \includegraphics[width=5.2cm]{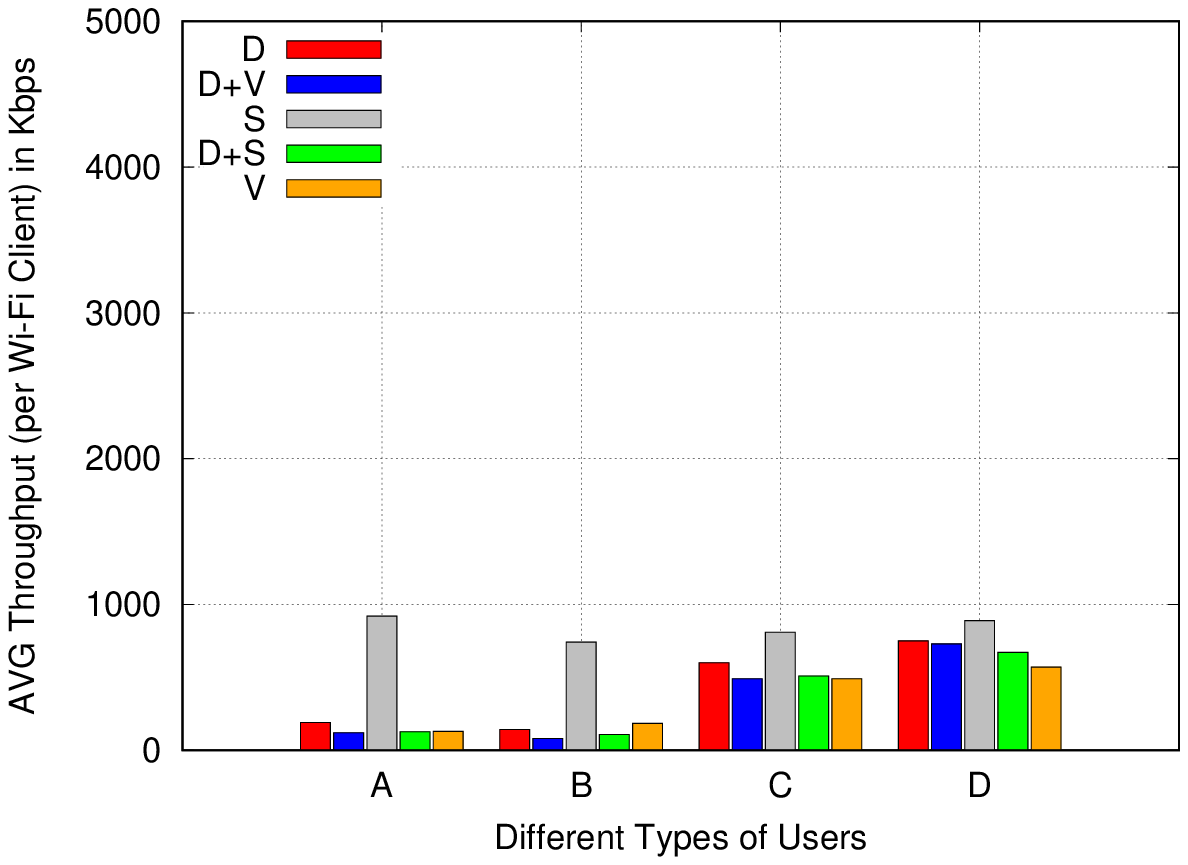}
  \caption{Wi-Fi Tput on Wi-Fi/LAA}\label{fig:laawifi36}
\end{subfigure}
\caption{LAA and Wi-Fi Throughput, Wi-Fi clients associated to three Wi-Fi APs, Scenario S2}
\label{a1}
\end{figure*}

\subsubsection{Wi-Fi throughput when coexisting with LAA} 
Fig.~\ref{wifiTputW} (a), (b), and (c) show the baseline throughput when Wi-Fi coexists with itself in scenarios S1, S2 and S3, respectively and Fig.~\ref{wifiTputL} (a), (b), and (c) show the throughput when Wi-Fi coexists with LAA in scenarios S1, S2 and S3, respectively.
The lower throughput on Wi-Fi client D (Xiaomi phone) in these figures is due to this device supporting only 802.11n. 
Comparing the two sets of figures, we observe that the average throughput per Wi-Fi client when Wi-Fi coexists with itself (Fig.~\ref{wifiTputW}, Wi-Fi/Wi-Fi) is much higher than when it coexists with LAA (Fig.~\ref{wifiTputL}, Wi-Fi/LAA). The only traffic type that does not experience a degradation in throughput is real-time streaming (S) and this is because LAA does not use the unlicensed band at all for this traffic type, as we saw previously. In Scenario S2 were all the clients are close to the AP and in line-of-sight, the largest decrease in average throughput per client is 97\% when coexisting with LAA, in Scenario S1 the highest decrease of average throughput per client is 89.8\% and in Scenario S3, where the clients are distant from the Wi-Fi AP and the throughput is low even in the absence of LAA, the largest decrease of average throughput per client is 82.4\%. It should be noted that the Wi-Fi throughput in general is lower than LAA even when coexisting with itself: we verified that this is due to the University backhaul network for Wi-Fi being limited to approximately 65 Mbps, unlike the LAA BS which is not controlled by the University.

\subsection{Throughput when all three APs are used in Scenario S2}

Since all APs were not reachable at all channels and locations, we present results only for Scenario S2 where all APs were being received with good signal strength. When Wi-Fi/Wi-Fi coexistence is being studied, the five Wi-Fi clients are associated as follows: clients A and B to AP2, clients C and E to AP1, and client D to AP3. When Wi-Fi/LAA coexistence is being studied, client E is used as the LAA client.

Fig.~\ref{a1} (b) and (c) show Wi-Fi throughput with Wi-Fi/Wi-Fi and Wi-Fi/LAA respectively. As in the previous experiment, we again observe a reduction in Wi-Fi throughput when Wi-Fi coexists with LAA compared to the Wi-Fi/Wi-Fi scenario, with the largest decrease in average throughput per client being 85.1\% when coexisting with LAA.
We also see in Fig.~\ref{a1} (c) a slightly higher throughput on client C which is associated with AP1 and client D which is associated with AP3, compared to clients A and B which are both associated with AP2: this is because AP2 has two clients associated with it. We also note that AP2 uses Channel 157 as its primary channel and since there are two clients associated with it, management traffic is higher and there is a corresponding lower throughput on Channel 157 for the LAA client as seen in Fig.~\ref{a1} (a).
Interestingly, we see that when the Wi-Fi clients are distributed over all three Wi-Fi APs, LAA can transmit on all three unlicensed channels, as shown in Fig.~\ref{a1} (a), unlike the previous experiment where all Wi-Fi clients were connected to one Wi-Fi AP and only two unlicensed channels were being used by LAA as shown in Fig.~\ref{atla} (b). Hence the largest decrease of aggregate throughput from the baseline is only 13.8\% compared to the 62\% reduction when all Wi-Fi clients were connected to one AP.

\section{Conclusions and future work}

The experimental scenario we have measured in this paper is one that is of considerable interest and not been studied in the coexistence literature to date. The conventional wisdom is that if Wi-Fi is deployed indoors and LAA is deployed outdoors, there will be minimal degradation in performance. However, we have demonstrated that in a realistic environment, like an university campus, where Wi-Fi usage is considerable in busy outdoor areas such as outside a bookstore, the presence of an outdoor LAA BS in close proximity can significantly affect Wi-Fi performance: Wi-Fi throughput can degrade as much as 97\% when coexisting with LAA whereas LAA throughput only degrades 35\%. Furthermore, we have made a number of interesting conclusions regarding coexistence between Wi-Fi and LAA under such circumstances, which are not apparent from existing analysis and simulations. For instance, when both Wi-Fi and LAA use multiple channels, the choice of primary channels among the different APs can have a substantial effect on the throughput of LAA due to the increase in Wi-Fi management frames like RTS/CTS. This leads to the surprising result that LAA has higher throughput when coexisting with three 80 MHz Wi-Fi APs operating on the same channel but using different primary channels than with one 80 MHz Wi-Fi AP when the total number of clients in both cases is the same. 

Since the work described in  this paper relied on deployed LAA and off-the shelf Wi-Fi APs, we were unable to change parameters like sensing thresholds and TXOP duration's, or implement preamble detection instead of energy detection as a coexistence mechanism, to determine the effects on performance. Our future work will focus on addressing the unique problems we identified in this paper via analysis, system simulations using ns-3, and SDR testbeds.

\end{document}